\documentclass{article}
\usepackage{PRIMEarxiv}
\usepackage[utf8]{inputenc}
\usepackage[T1]{fontenc}
\usepackage{hyperref}
\usepackage{url}
\usepackage{booktabs}
\usepackage{amsmath,amssymb,amsfonts,bm}
\usepackage{nicefrac}
\usepackage{microtype}
\usepackage{fancyhdr}
\usepackage{graphicx}
\usepackage{algorithm}
\usepackage{algorithmic}
\graphicspath{{images/}}

\def\beq{\begin{eqnarray}}
\def\eeq{\end{eqnarray}}
\def\be{\begin{EQA}[c]}
\def\ee{\end{EQA}}

\def\bel{\begin{EQA}[c] \label}

\newcommand \bei {\begin{itemize}}
\newcommand \eei {\end{itemize}}

\def\beel{\begin{eqnarray} \label}

\DeclareMathOperator{\diag}{diag}

\title{Quantum-enhanced policy iteration \\ on the example of a mountain car}

\author{
  Egor E. Nuzhin \\
  Skolkovo Institute of Science and Technology \\
  Moscow 121205, Russia \\
  \texttt{e.nuzhin@skoltech.ru} \\
  \And
  Dmitry Yudin \\
  Skolkovo Institute of Science and Technology \\
  Moscow 121205, Russia \\
}

\begin{document}
\maketitle

\begin{abstract}
Advances in the experimental demonstration of quantum processors have provoked a surge of interest in the idea of practical implementation of quantum computing over last years. It is expected that the use of quantum algorithms will significantly speed up the solution to certain problems in numerical optimization and machine learning. In this paper, we propose a quantum-enhanced policy iteration (QEPI) algorithm as widely used in the domain of reinforcement learning and validate it with the focus on the mountain car problem. In practice, we elaborate on the soft version of the value iteration algorithm, which is beneficial for policy interpretation, and discuss the stochastic discretization technique in the context of continuous state reinforcement learning problems for the purposes of QEPI. The complexity of the algorithm is analyzed for dense and (typical) sparse cases. Numerical results on the example of a mountain car with the use of a quantum emulator {and available quantum hardware as provided by D-Wave} verify the developed procedures and benchmark the QEPI performance.   
\end{abstract}
\maketitle

\section{Introduction}

{A plethora} of existing methods of numerical optimization has been greatly enriched over last years through the advent of first prototypes of quantum processors~\cite{Harrigan2021,Ebadi2022,Yarkoni2022,Nguen2023}. Remarkably enough, a lot of the methods developed are based on quantum-inspired classical algorithms, which are in many cases capable of solving numerical optimization and machine learning {tasks} much more efficiently than their traditional counterparts. A particular example is quantum optimization as based on the approach known as adiabatic quantum computing~\cite{Farhi2000}. In this case, the optimization problem to be addressed is associated with a certain physical system specified by the Hamiltonian, which exactly coincides with the cost function of the optimization problem. Thus, finding the ground state of this Hamiltonian gives an optimal solution to the initial problem. {Technically}, this procedure is implemented as follows. The algorithm begins by considering a system described by some {\it simple} Hamiltonian, chosen in such a way that its ground state is known and can be easily prepared. We adiabatically deform the {\it simple} Hamiltonian so that, after a long time, we obtain the Hamiltonian of the original problem. This procedure has a physical justification in the framework of the adiabatic theorem, according to which a system in the ground state will always remain close to its instantaneous ground state, provided that its lower energy levels are nondegenerate and evolution proceeds rather slowly~\cite{Born1928}. In practice, the time of adiabatic evolution is usually chosen inversely proportional to the square of the minimum energy difference between the instantaneous ground state and the first excited state. This is an example of a universal model of quantum computing, which is capable of emulating any quantum algorithm.

The most remarkable example is the so-called quantum annealing {that can be implemented with the use of real devices}~\cite{Finnila1994,Kadowaki1998,Das2008}. This process is similar to the classical simulated annealing algorithm~\cite{Pincus1970,Khachaturyan1979,Kirkpatrick1983}. In simulated annealing algorithm, the transition between local minima on the manifold specified by a given Hamiltonian is attributed to thermal fluctuations. With decreasing temperature, the probability of a transition to the global minimum increases significantly. During quantum annealing, the transitions between local minima are due to quantum tunneling. This process makes it possible to study the variety of local minima much more efficiently compared to thermal noise, especially when the energy barriers are sufficiently high and narrow. The quantum annealing approach is ideal for solving NP-hard combinatorial optimization problems, including the traveling salesman problem. In this paper, we show how one can benefit from the use of quantum annealers in the domain of reinforcement learning (RL) that is widely considered as one of the machine learning paradigms~\cite{Li2017,Mousavi2018}. 
{It is worth mentioning previous studies highlighting quantum RL (QRL) algorithms, being mostly devised from the perspective of variational quantum computing. A particular example~--~quantum policy iteration via Grover search (QPI-GS) \cite{wiedemann2022quantum}~--~is known to utilize the amplitude estimation to assess the value function and Grover search to improve the policy. This algorithm is used for finite horizon RL tasks with its complexity being quite sensitive to the horizon length. The other algorithm~--~QRL via policy iteration (QRL-PI)~\cite{cherrat2023quantum}~--~was developed for infinite horizon problems and encodes the value function in a similar fashion, {\it i.e.}, in the form of the amplitude. In QRL-PI, one extensively employs quantum linear system solvers using block-encoding~\cite{chakraborty2018power}. The solvers appear as an extension of fundamental HHL \cite{harrow2009quantum,duan2020survey} algorithm. In contrast to these approaches, we suggest applying quantum annealing as a subroutine to find the solution to a system of linear equations (SLE)~\cite{chang2019quantum}.}

Let us briefly recall standard iteration algorithms that are of wide use in RL, namely value iteration (VI) and policy iteration (PI)~\cite{howard1960dynamic}. In the former case, the algorithm starts with an arbitrarily chosen value function $V(s)$ with the follow-up update according to the equation,
\begin{equation}
\label{vi}
    V(s) =\max_a \sum_{s',r} p(s',r|s,a)\cdot\{r + \gamma V(s')\},
\end{equation}
where \(s\) stands for the current agent state and \(a\) is the agent action applied in the state \(s\), whilst \(s'\) is the state next after the agent transition from state \(s\), which is defined by transition probability \(p(s',r|s,a)\). It should also be mentioned that $r$ is the reward generated by taking the action $a$ and $\gamma$ is the discount factor for future rewards. In~\eqref{vi}, we average over all possible $r$ and $s'$. The policy is then restored by using a one-step look-ahead:
\begin{equation}
    \pi(s) = \arg \max_a \sum_{s',r} p(s',r|s,a)\cdot\{r + \gamma V(s')\}.
\end{equation}
In the case of PI, one makes use of the same equations with policy evaluation stage being implemented first. The policy-conditioned value function is updated until convergence based on
\begin{equation}
\label{pi_v}
    V(s) = \sum_{s',r} p(s',r|s,\pi(s))\cdot\{r + \gamma V(s')\}.
\end{equation}
As opposed to a full probabilistic description, it is advisable to directly employ functional transition \(s' = s'(s,a)\) in some cases with the value function update being implemented as
\begin{equation}
\label{pi_v_det}
    V(s) = r(s,\pi(s),s'_\pi) + \gamma V(s'_\pi),
\end{equation}
where \(s'_\pi = s'(s,\pi(s))\). When convergence is achieved, the policy is improved according to
\begin{equation}
\label{pi_pi}
    \pi(s) = \arg \max_a\sum_{s',r} p(s',r|s,a)\cdot\{r + \gamma V(s')\}.
\end{equation}
The algorithm is repeated until a stable policy is attained. In the following, we develop the soft version of the VI algorithm, which is suitable for policy interpretation and provides intuitive understanding, and discuss in detail quantum-enhanced policy iteration (QEPI).

\section{Problem statement}
\begin{figure}[htbp]
    \centering
    \includegraphics[width=6cm]{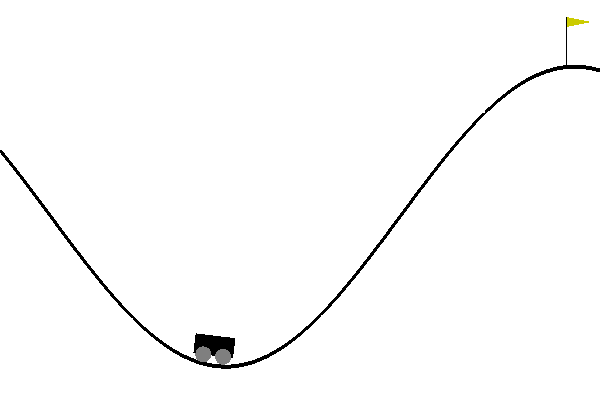}
    \caption{A schematic of the mountain car environment. In this scenario, the agent is the car in the valley. The agent aims at reaching the flag on top of the hill.}
  \label{fig:mc}
\end{figure}

We demonstrate the application of the method on a mountain car problem. Although being very basic and simple it allows to illustrate the main ideas without shading them by technical details. The problem of a mountain car is a well-established testing framework in the domain of RL~\cite{moore1990efficient}. In this scenario, a car moves upwards or downwards a hill depending on action policy. The goal is to reach the top of the hill. Hence the reward is negative until it passes through the top of the hill and the reward is zero otherwise. The set of actions is predefined and discrete, namely the car might accelerate to the left or to the right, or has no acceleration, as shown in Fig.~\ref{fig:mc}. Note that the car engine is too weak to overcome the pulling down gravitational force on the hill; that is, a simple policy of always driving to the right is not applicable. The state space consists of the car's speed and position with both of them being continuous variables. In the following, we choose the agent transition to obey the differential equation
\begin{equation}
    \ddot{x} =  (a - 1) \cdot f - g \cdot \cos(3x),
\end{equation}
that represents the trade-off between the engine's thrust and gravity on the car~\cite{brockman2016openai}. Here, \(x\) is the position of the car, \(a \in\{0,1,2\}\) specifies a possible action, whilst \(f\) and \(g\) are force and gravity constants, respectively. 

\begin{figure}[htbp]
    \centering
    \includegraphics[width=5cm]{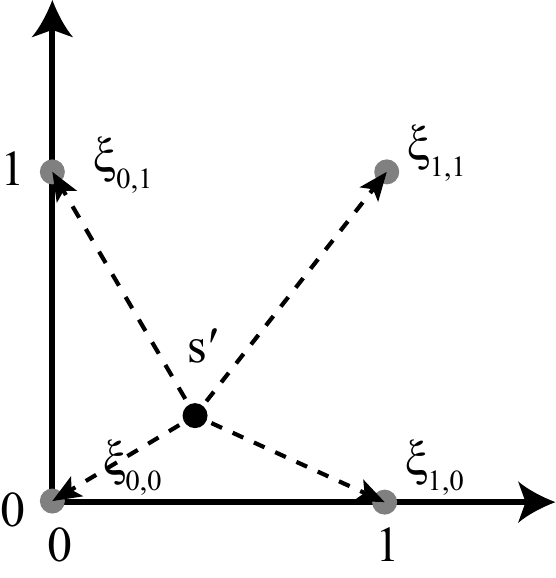}
    \caption{A schematic representation of stochastic transition in the form of random walks from the state $s^\prime$ to neighboring vertices $\xi_{00}$, $\xi_{10}$, $\xi_{01}$, and $\xi_{11}$, positioned in the corners of a unit square. The coordinates are normalized to unity.}
  \label{fig:st_on_nv}
\end{figure}

As a rule of thumb, both VI and PI algorithms are discussed in the context of discrete Markov decision processes~\cite{guo2009continuous}, whereas the position and velocity of a mountain car are continuous in nature. This can be addressed by virtue of discretization. In the following, we will be dealing with stochastic transitions which are defined by hopping from a given vertex to one of the neighboring vertices \cite{kushner1990numerical,hinze2012discretization}, that is known to be a simplified version of the Kuhn triangulation~\cite{munos2002variable}. In our case the state manifold corresponds to a two dimensional grid, with the respective transition probabilities. That is, a transition from state \(s\) to state \(s'\) implies a random walk to the one out of four nearest vertices of the mesh, which may be denoted as \(\xi_{10}\), \(\xi_{11}\), \(\xi_{00}\), and \(\xi_{01}\) with the probability that is proportional to the distance between these points. The transition probabilities are listed in Table \ref{table:st_on_nv}, and the coordinates on the grid are normalized so that the grid points are set on corners of the unit cell with the axes origin in one of the corners. The hopping to a neighboring vertex, depicted in Fig.~\ref{fig:st_on_nv}, is easy to implement so that the acceptable tolerance on the coarse mesh is achieved. It is valuable enough as a practically available quantum computer does not contain many qubits. However, a fine mesh may require significant computational resources, and a simple deterministic snap onto the nearest vertex often results in transitions to the same grid point on a coarse mesh, which is not an acceptable system dynamics approximation.

\begin{table}[htbp]
\centering
\caption{Probabilities of hopping from a given vertex to one of the
neighboring vertices. Coordinates are normalized to unity.}
\label{table:st_on_nv}
\begin{tabular}{lcccc}
\toprule
State       & $\xi_{00}$    & $\xi_{01}$ & $\xi_{10}$ & $\xi_{11}$ \\
\midrule
Probability & $(1-x)(1-y)$  & $(1-x)y$   & $x(1-y)$   & $xy$ \\
\bottomrule
\end{tabular}
\end{table}

The Bellman equation~\eqref{pi_v_det} for the state value function~\cite{bellman1957markovian} can be rewritten in this case as
\begin{equation}
    V(s) = \sum_{s'} p(s'|s,\pi(s))\cdot\{r(s') + \gamma V(s')\}
\end{equation}
with \(s\) and \(s'\) being specified on the grid and \(p(s'|s,a)\) being non-zero for neighboring to \(s'(s,a)\) grid nodes only. It is worth mentioning that the value iteration algorithm, applied to the continuous state environment, does not produce smooth results in terms of the final optimal policy and value function as visualized in Figs.~\ref{fig:policies_base} and \ref{fig:values_base} (VI). Generally speaking, the resulting solution is unlikely to be interpreted and analyzed with the problem being however relaxed by adopting noise-assisted smoothing. This noise is defined as a Gaussian deviation on the agent's departure state, {\it i.e.},  
\begin{equation}
    \tilde{s} \sim \mathcal{N}(\tilde{s}\,|\,s,\sigma^2)
\end{equation}
where \(\tilde{s}\) is the perturbed agent state and \(\mathcal{N}\) is the multivariate normal distribution with the mean value $s$ and variance $\sigma^2$.

Substituting the transition probability in Eq.~\eqref{vi} for the value function update we arrive at  the multivariate convolution with the Gaussian kernel,
\begin{equation}
    V(s) = \max_a \, \left[\mathcal{N}(\cdot|0,\sigma^2) * Q(\cdot,a)\right](s),
\end{equation}
where
\begin{equation}
    Q(s,a) = \sum_{s',r}  p(s',r|s,a)\cdot\{r + \gamma V(s')\}.
\end{equation}
The convolution is implemented in a range of packages and is also known as the Gaussian blur~\cite{gedraite2011investigation} adopted in image denoising. We are to apply the blur to find a smooth solution to the mountain car problem and compare it with that obtained with a quantum annealer. Remarkably, smoothing technique as implemented this way may be considered as a tool to find a robust solution to the RL problem, in which precise transition of the agent according to model dynamics is not possible. This imprecise transition may be a meaningful feature of RL agents in real systems tending to demonstrate trembling hand policies~\cite{hansen2010computational}.

\section{Quantum-enhanced policy iteration}

We herein analyze the use of a quantum annealer for the RL task as specified above. A close inspection of the policy iteration algorithm allows one to split it into two sequential steps, namely policy evaluation and policy improvement, which are repeated until convergence. The result of the first stage is the value function \(V(s)\) that satisfies the relation
\begin{equation}
     V(s) = \sum_{s',r} p(s',r|s,\pi(s))\cdot\{r + \gamma V(s')\}.
\end{equation}
This equation can be easily linked to the solution of SLE in the form $Ax = b$, where \(x\) is vector corresponding to value function such as \(x_i = V(s_i)\), while
\begin{equation}
\label{b}
    b_i = \sum_{s',r} p(s',r|s_i,\pi(s_i)) \cdot r,
\end{equation}
and the matrix $A$,
\begin{equation}
\label{A}
    A_{ij} = \delta_{ij} - \delta_{s_j \not\in \, \mathrm{term}} \cdot\gamma \cdot \sum_{r} p(s_j,r|s_i,\pi(s_i)) .
\end{equation}
Note that in~\eqref{A} we have respected the terminal conditions, which are prevalent in RL, supplementing the last term with \(\delta_{s_j \not\in \,\mathrm{term}}\) that is zero for terminal states.

Finding the solution to SLE, $x^\ast$, is equivalent to the policy evaluation stage in the PI algorithm. We are using binary encoding to transform the SLE into a quadratic unconstrained binary optimization (QUBO) problem, {\it i.e.}, for each state we introduce \(n_b\)-dimensional binary variable \(y_i \in \{0,1\}\) provided that
\begin{equation}
\label{bin_enc}
    x_i = -\kappa \cdot\sum_{j=0}^{n_b-1} 2^j\cdot y_{i+\mu \cdot j}, \quad \kappa=\frac{|x_\mathrm{min}|}{2^{n_b}-1},
\end{equation}
where \(\mu\) is the total number of states, \(x_\mathrm{min}\) is the minimal possible value of variable \(x\) and \(x_i\) stands for the value function at \(i\)-th state. Note that \(x\) is negative since the negative reward is typical for classical control problem such as mountain car, but the procedure may be generalized for arbitrary reward values. Clearly, 
\begin{equation}
    x^* = \arg \min_{x} ||A x - b||^2.
\end{equation}
In the following, we define the block matrix \(P\), such that the block of that matrix,
\begin{equation}
\label{P}
    P_{ij} =  2^{i+j}\cdot \kappa^2 \cdot A^TA,  
\end{equation}
and the vector,
\begin{equation}
\label{p}
    p_i = 2^{i+1}\cdot\kappa \cdot b^T A,
\end{equation}
for \(i, j = 0,\ldots,n_b-1\). Plugging \(x\) rewritten in terms of \(y\) as specified by~\eqref{bin_enc} into the optimization task allows to reduce the problem to
\begin{equation}
\label{obj}
   y^* = \arg \min_{y} (y^TPy + p^Ty).
\end{equation}

When the binary problem is solved, the solution may be restored by summing up corresponding weighted variables using relation~\eqref{bin_enc}. One can clearly notice that QEPI is somewhat similar to PI with a minor difference in that we translate the policy evaluation stage to a quantum computer. We provide Algorithm \ref{alg:qpi} below for reference. 

\begin{algorithm}
\caption{Quantum-enhanced policy iteration}
\label{alg:qpi}
\textbf{Input}: initial value function: \(x\); initial policy: \(\pi\); discount factor \(\gamma\).
\begin{algorithmic}[1]
\FOR{$k = 0,1,2,\ldots$}
\STATE  Build SLE correspondent to a particular RL problem to evaluate current policy:
\begin{equation}
\nonumber
 A_{ij} = \delta_{ij} -   \delta_{s_j \not\in \, \mathrm{term}} \cdot\gamma \cdot \sum_{r} p(s_j,r|s_i,\pi(s_i))
 \end{equation}
\begin{equation}
\nonumber
    b_i = \sum_{s',r} p(s',r|s_i,\pi(s_i)) \cdot r
\end{equation}
\STATE Transform SLE to QUBO:
\begin{equation}
\nonumber
   y^\ast =  \arg \min_{y} (y^TPy+p^Ty).
\end{equation}
\STATE Solve the resulting QUBO problem with a given quantum annealer or its simulator.
\STATE Restore value function corresponding to current policy from QUBO solution.
\STATE Update the policy:
\begin{equation}
\nonumber
    \pi(s) = \arg \max_a\sum_{s',r} p(s',r|s,a)\cdot\{r + \gamma V(s')\}
\end{equation}
\ENDFOR
\end{algorithmic}
\end{algorithm}

\section{Computational complexity}
\label{sec:procedure}
The complexity of the algorithm depends on the specifics of the RL problem. Assume that the transition tensor \(p(s',r|s,a)\) is dense, {\it i.e.}, the number of zero entities in the tensor is sufficiently small. In this case, time complexity of the algorithm might be estimated as
\begin{equation}
    T_\mathrm{dense} = \mathcal{O}(\mu^3 + \left(n_b^2+\alpha\rho\right)\mu^2 + \tau_\mathrm{QA}),
\end{equation}
where $\rho$ is the maximal number of distinct rewards possible in conditional (on \(s,a,s'\)) transition, and $\alpha$ is the total number of agent actions. The value \(\tau_\mathrm{QA}\) describes the time complexity of annealing operation in a quantum computer, which is a complex problem-specific function~\cite{morita2008mathematical}. The time complexity derivation is discussed in detail in Appendix~\ref{sec:dense_time_complexity}. Similarly, space complexity elaborated in Appendix~\ref{sec:dense_space_complexity} in the case of dense transition tensor, yields
\begin{equation}
        M_\mathrm{dense} = \mathcal{O}((n_b^2 + \alpha\rho)\cdot\mu^2).
\end{equation}
\begin{figure}[htbp]
    \centering
    \includegraphics[width=\textwidth]{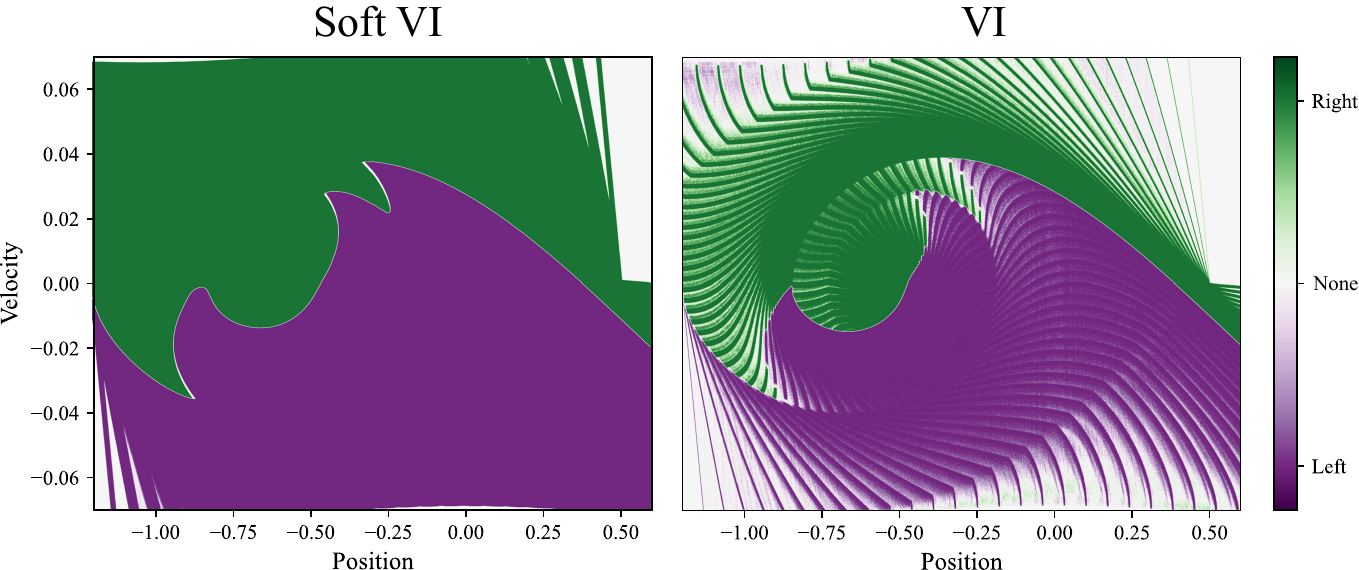}
         \caption{Optimal policies corresponding to value iteration (VI) and soft value iteration (Soft VI) in terms of the phase space as specified by position and velocity of the mountain car. The algorithms were performed for 400 timesteps on a grid $6000\times 6000$ units and $\gamma = 0.99$. In case of Soft VI, smoothing is chosen $\sigma = 10$ of the grid size.}
         \label{fig:policies_base}
\end{figure}

Meanwhile, typical problems in RL have a sparse structure of transition probability function. This structure is caused by state space continuity with forbidden instantaneous long jumps between the states. In this case, the transition probability tensor has a band structure, explained in Appendix~\ref{sec:band_tensors}, with respect to state indices \(s\) and \(s'\). If the transition is not allowed between states \(s\) and \(s'\), so that \(||s-s'||_\infty>k\), we call such transition tensor band with the bandwidth \(k\).
Provided band transition probability time complexity corresponds to
\begin{equation}
    T_\mathrm{sparse} = \mathcal{O}\left(\mu ac+\mu cn_b^2+ \alpha\rho\mu a + \tau_\mathrm{QA}\right),
\end{equation}
where $a=(2k+1)^N$ and $c=(4k+1)^N$ in the \(N\)-dimensional state space, as detailed in Appendix~\ref{sec:sparse_time_complexity}. It is by two orders lower than that for the dense transition probability with respect to the total number of states \(\mu\). And for the space complexity in the case of sparse transition matrix we have
\begin{equation}
    M_\mathrm{sparse} = \mathcal{O}\left(\left(\alpha\rho a+cn_b^2\right)\cdot\mu\right),
\end{equation}
which is by one order in \(\mu\) lower than the dense counterpart as discussed in Appendix~\ref{sec:sparse_space_complexity}.

As to quantum resources, the QEPI algorithm necessitates a quantum annealer with no less than
\begin{equation}
\label{nqubits}
    N_\mathrm{qubits} = \mu n_b
\end{equation}
qubits, since each binary variable of the QUBO problem is mapped onto a single qubit,
provided qubits' topology~\cite{shin2014quantum} allows encoding a vectorized band
tensor of the bandwidth \(k\) and corresponding sparsity, {\it i.e.}, fraction of zero
entities,
\begin{equation}
    \textrm{Sparsity}(Q) \ge 1-\frac{c}{\mu},
\end{equation}
where \(Q = P + \diag(p)\) is introduced to describe the binary quadratic form in Eq. \eqref{obj} with a single matrix.
For a rigorous derivation of this result we refer to Appendix~\ref{sec:qa_requirenments}.

\section{Results}

We herein describe the results as provided by VI and the proposed soft VI as well as the introduced QEPI applied to the mountain car problem. The problem has been discretized following the idea of the stochastic transition to nearest neighbors, so that the VI algorithm can be easily adopted for coarse meshes. Fine mesh requires a significant amount of memory for sufficient problem binarization. The soft VI algorithm was proposed to get an interpretive picture of the action policy, which we compare with the solution as obtained based on quantum-enhanced PI. 

\begin{figure}[htbp]
    \centering
    \includegraphics[width=\textwidth]{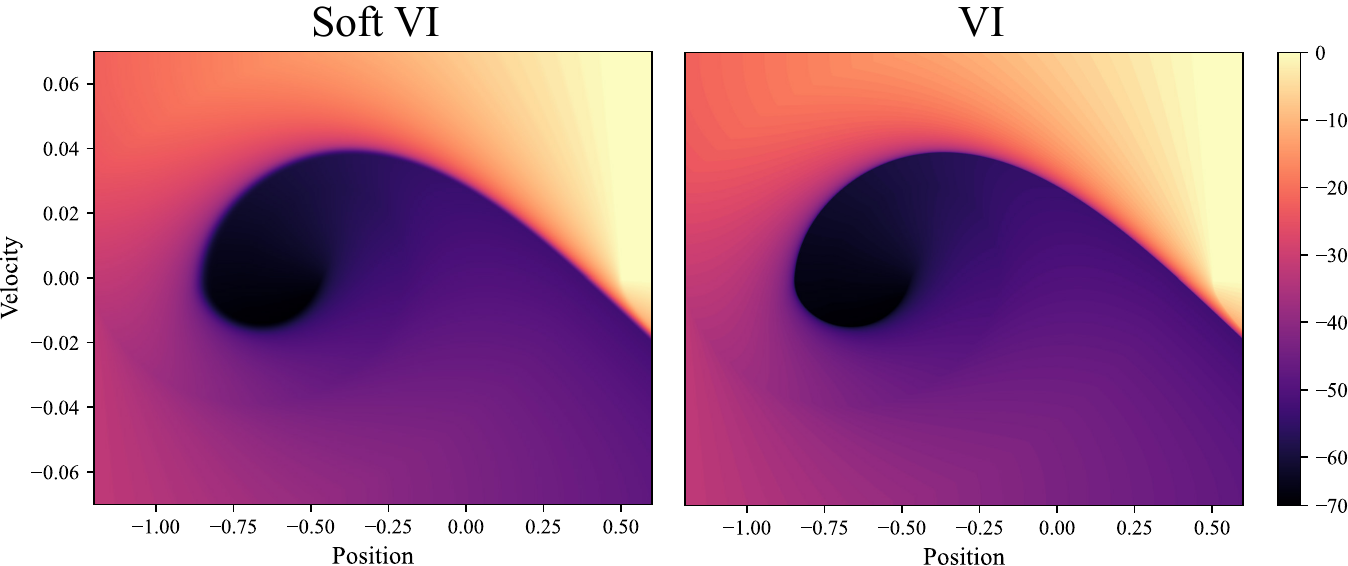}
    \caption{Optimal value functions corresponding to value iteration (VI) and soft value iteration (Soft VI). The algorithms were performed for 400 timesteps on a grid $6000\times 6000$ units and $\gamma = 0.99$. In case of Soft VI, smoothing is chosen $\sigma = 10$ of the grid size.}
    \label{fig:values_base}
\end{figure}

\begin{figure}[htbp]
    \centering
    \includegraphics[width=\textwidth]{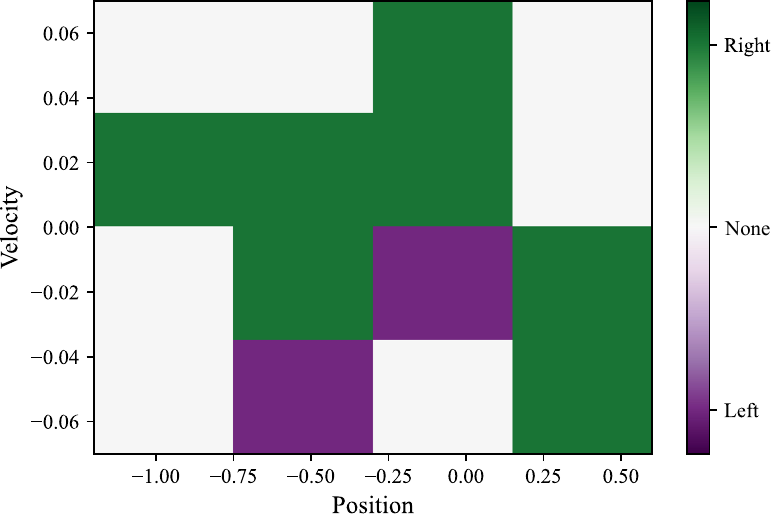}
    \caption{Optimal policy corresponding to Quantum-enhanced policy iteration (QEPI) and value iteration (VI). The algorithms were performed for 10 policy update steps on a grid $4\times4$ units and $\gamma = 0.99$. For QEPI we apply number of bits $n_b = 10$, $x_\mathrm{min} = -100$, and the number of anneals per policy update is 100.}
         \label{fig:poilcy_qpi}
\end{figure}

One can clearly notice the difference between VI and soft VI action policies on the mountain car problem in Fig.~\ref{fig:policies_base}. Obviously, the soft VI policy is much simpler, having fewer areas with distinct action strategies and simpler shapes. Value functions as visualized in Fig.~\ref{fig:values_base} are quite similar, but the soft VI value function has smoother edges, which is expected since it is retrieved with the Gaussian blur, which is used in image denoising. Absolute values of both functions have comparable scales inducing a conclusion that smoothing has a minor effect on the policy performance. Quantum-enhanced PI was applied on a coarse mesh. We validated the solution of QEPI on the solution of base VI with stochastic discretization on the nearest neighbor. The final optimal policy is depicted in Fig.~\ref{fig:poilcy_qpi}. In comparison to the policy generated by the soft VI algorithm, we can note similar areas corresponding to swinging motion and to no acceleration.

\begin{figure}[htbp]
    \centering
    \includegraphics[width=\textwidth]{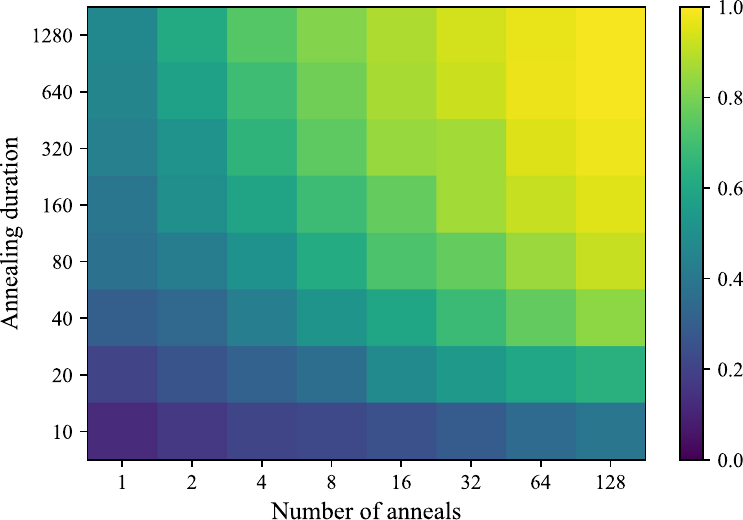}
    \caption{Accuracy of finding the optimal policy. Training curves are depicted for a range of annealing duration and number of anneals parameters. Accuracy is estimated in 1000 algorithm runs at the 10-th policy update step.}
         \label{fig:training_qpi}
\end{figure}

\begin{figure}[htbp]
    \centering
    \includegraphics[width=\textwidth]{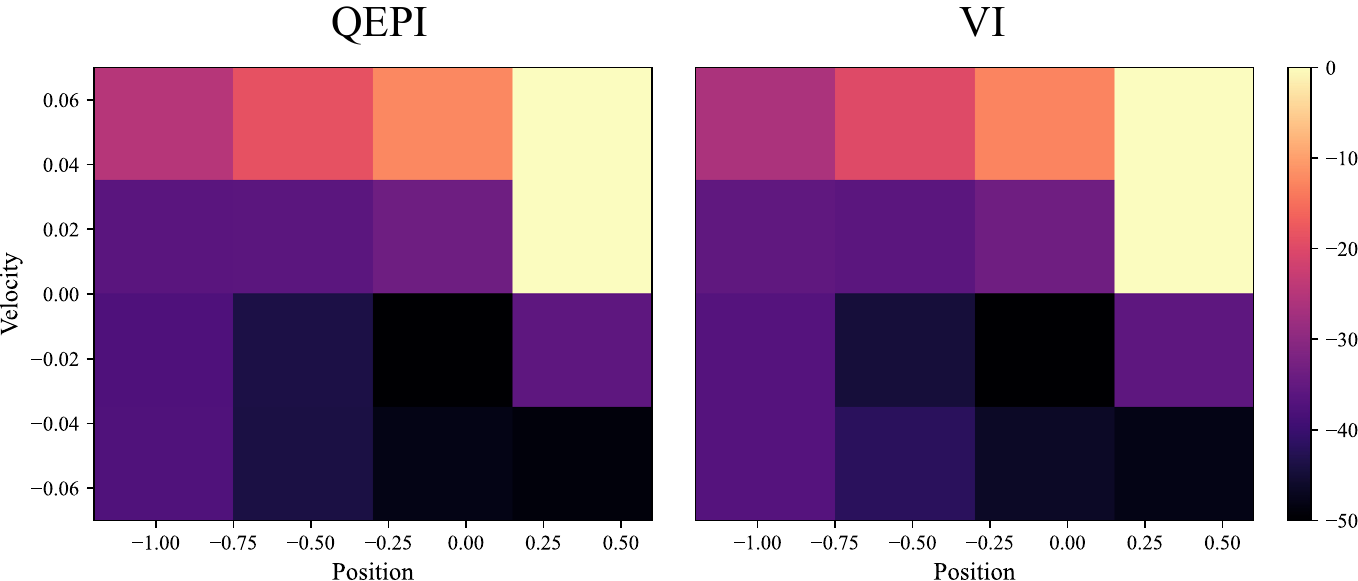}
    \caption{Optimal value functions corresponding to quantum-enhanced policy iteration (QEPI) and value iteration (VI). QEPI was performed for 1280 annealing duration steps.}
    \label{fig:values_qpi}
\end{figure}

We performed the accuracy estimate for the QEPI algorithm. We put as an ideal solution that obtained with the VI algorithm and considered QEPI successful if the policy produced by QEPI coincided with the value iteration policy over the entire state range without exceptions. Fig.~\ref{fig:training_qpi} allows one to visualize the dependence of QEPI accuracy versus number of sequential anneals  and annealing duration steps for a fixed number of policy update steps, namely 10. The training curves validate that sufficient annealing duration and a number of anneals parameters ensure the convergence of the QEPI algorithm to the optimal policy. In Fig.~\ref{fig:values_qpi}, we compare values functions as obtained with the QEPI and VI algorithms. The functions have almost identical values, and a little difference here is a consequence of binarization. Due to binarization, the range of function values is limited to a set of predefined levels. The solution of QEPI approaches the one delivered by VI when the accuracy of binarization on a quantum annealer approaches the accuracy of discretization on a classical computer. Note that the QEPI algorithm was implemented with PyTorch~\cite{Pytorch} for tensorial operation on GPU and Qubovert~\cite{Qubovert} package for quantum annealing simulation.

\begin{figure}[htbp]
    \centering
    \includegraphics[width=\textwidth]{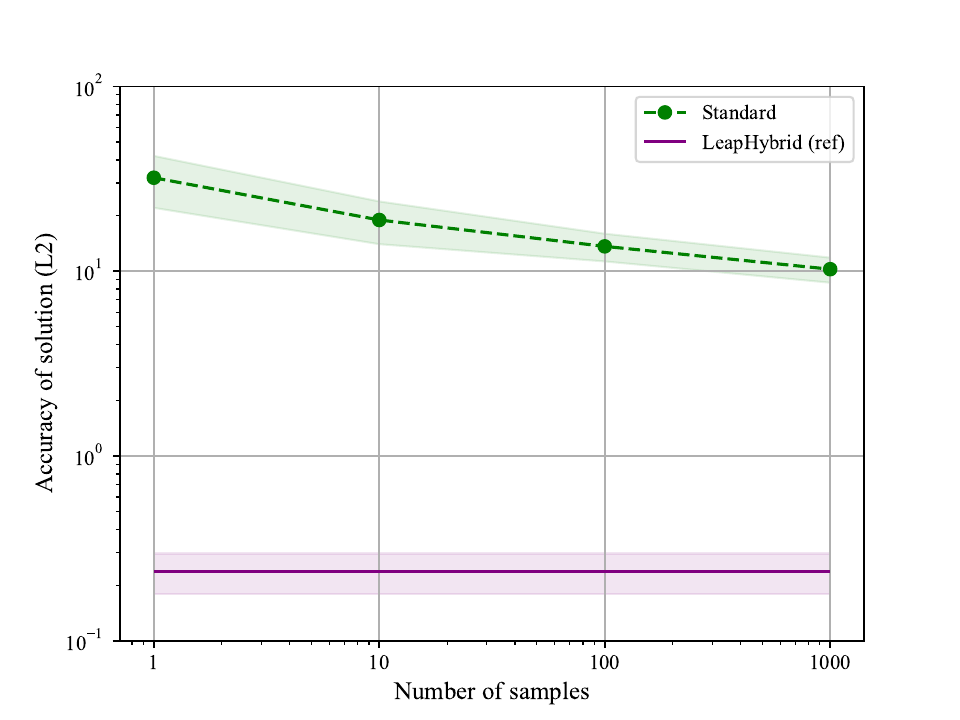}
    \caption{{Accuracy of SLE solution  (the first iteration in QEPI) depending on the number of samples for the standard D-Wave sampler. The accuracy was estimated on 100 evaluations. The accuracy of the Leap's hybrid solver is shown for reference. Shadow areas represent standard deviation.}}
         \label{fig:accuracy_dwave_samples}
\end{figure}

\begin{figure}[htbp]
    \centering
    \includegraphics[width=\textwidth]{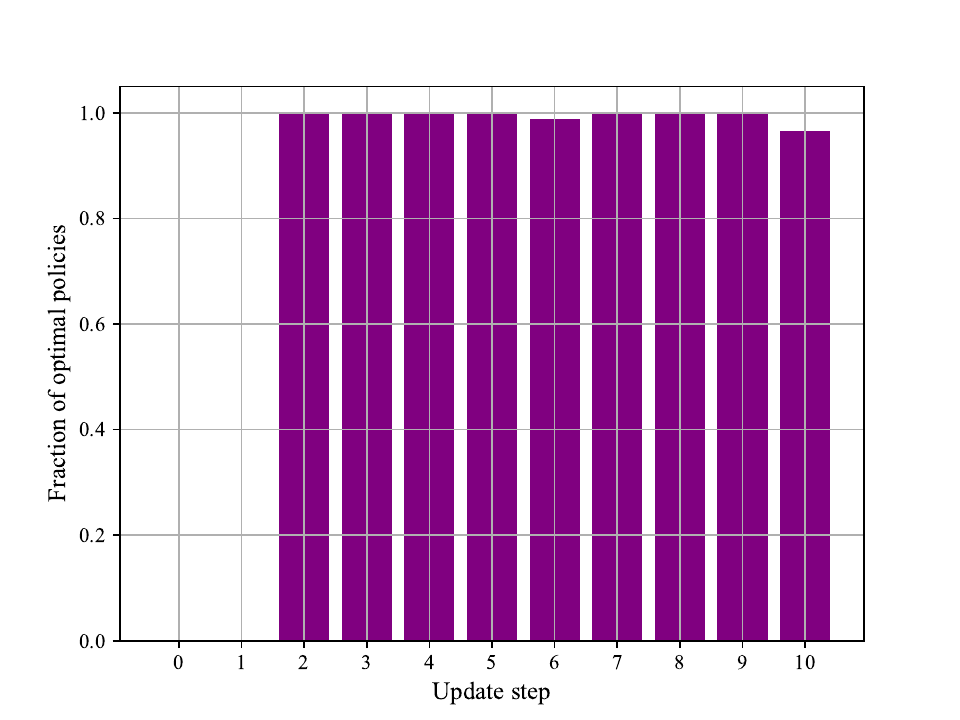}
    \caption{{Accuracy of finding the optimal policy depending on policy update step. The QEPI algorithm is implemented using a piece of quantum hardware as provided by D-Wave with the Leap's hybrid algorithm to solve the SLE. The fraction of optimal policies was estimated on 40 consequent algorithm runs.}}
         \label{fig:training_dwave}
\end{figure}

{To challenge the developed methodology we adapted the QEPI algorithm to a piece of available quantum hardware. Specifically, we carried out experimental studies with the use of a D-Wave quantum system~\cite{dwave} as provided by both the standard annealing and the Leap's hybrid solver~\cite{dwave-sytems}. In Fig. \ref{fig:accuracy_dwave_samples}, we illustrate our findings on the solution to SLE with respect to the number of samples, {\it i.e.}, anneals. Clearly, upon increasing the number of samples allows one to reduce the computational error. However, the Leap's hybrid algorithm, that combines both the classic and quantum
solvers, is dramatically superior to the standard annealing routine. It should be
stressed that the hybrid solver relies on proprietary classical heuristics along
with the quantum processor, so that its performance cannot be attributed to
quantum resources alone. We employ it here as a practical tool to demonstrate the
convergence of QEPI at the algorithmic level rather than as evidence of a quantum
speed-up. The Leap's hybrid algorithm was applied to estimate the algorithm
convergence to the optimal policy. Fig.~\ref{fig:training_dwave} unambiguously reveals that the QEPI converges to the
optimal policy rather fast provided acceptable tolerance. Note that the experiments
were carried out on a coarse \(4\times4\) mesh, which corresponds to
\(\mu n_b = 160\) binary variables according to Eq.~\eqref{nqubits}. The linear
growth of the qubit requirement with the number of states is the principal
limitation of the approach on present-day hardware: a \(50\times50\) mesh with the
same binarization precision would already demand \(2.5\times10^4\) qubits, which
exceeds the capacity of currently available annealers. On the mesh addressed here
the underlying SLE is, of course, solved classically at a negligible cost, and the
purpose of the experiment is to validate the procedure rather than to demonstrate
a computational advantage.}

\section{Discussion and conclusion}

We proposed an algorithm to solve RL problems using available quantum processors. Particularly, we suggest using a quantum annealer in the context of the PI algorithm. In our approach, the policy evaluation stage is done with a quantum processor, whereas policy update step is performed on a classical computer. The policy evaluation stage is considered as a problem of solving SLE corresponding to policy-specific value function. To translate the continuous state RL problem we discuss using stochastic discretization techniques such as random hoppings to nearest neighbors. And to get an interpretable policy, we suggest using the proposed easy-to-implement soft value iteration. {The QEPI~--~Algorithm~\ref{alg:qpi}~--~is developed for  the general case of infinite horizon RL problems with a discrete state-action space. However, an arbitrary continuous state-action RL problem may be discretized, as was shown on the example of a mountain car problem, using stochastic transition to the nearest neighbor or with any other methods. Rewriting SLE in terms of QUBO~\cite{chang2019quantum} problem is not a specific task to RL either. However, it is important to choose a suitable binarization range and precision to cover the optimal solution of SLE with a given accuracy.}

We applied QEPI and soft VI to the mountain car problem. We discuss the advantage of soft VI to simplify optimal policy without significant losses of performance. In addition, the soft VI algorithm may be considered as a robust version of the environment model, in which exact transition is not possible, which makes sense in applied problems on real devices. The QEPI algorithm was tested using both the quantum annealer simulator and {quantum annealing system as provided by D-Wave.} Its convergence to the optimum was validated for a range of annealing parameters.

The algorithm may be easily applied to a range of RL tasks, especially the classical control problems, particularly the mountain car problem presented herein. It has linear classical time and space complexity with respect to state and action space for band (sparse) transition probabilities functions which are typical to a variety of problems in the domain of RL. The only quantum part of time complexity is an unknown variable, whose complexity properties are a subject of independent research. The QEPI algorithm is a promising candidate to accelerate the VI algorithm in the era of quantum computing, but it has disadvantages such as multiplicative growth of binary variables depending on the accuracy of binarization and complexity of fetching a global equilibrium in quantum annealing, which could require a number of sequential anneals and a long time of annealing duration. From another side, the sparse character of the linear operator is an advantage on existing hardware. The number of logical variables, \(\mu n_b\), is not affected by sparsity, but the connectivity requirements are relaxed: a logical variable whose degree exceeds that of the physical qubit graph has to be represented by a chain of several qubits. For a band \(Q\) the degree is bounded by \(c\) independently of \(\mu\), so that chains remain short.

{In comparison to classical RL algorithms such as classical VI we expect that the proposed methodology might be beneficial to speed up the policy evaluation step of the fundamental VI algorithm. Meanwhile, the policy gradient algorithms such as PPO~\cite{schulman2017proximal} can be still more efficient since they do not require an enormous amount of memory for the problem discretization containing the information in a condensed form of neural networks. It is worth mentioning that for infinite horizon problems QEPI does not need to trace the whole state-time sequence, in contrast to QPI-GS, whose qubit count grows linearly
with the horizon length. The qubit requirement of QEPI, Eq.~\eqref{nqubits}, is instead set by the number of states and the binarization precision. In comparison to QRL-PI algorithm the QEPI seeks the solution as a single sample on a quantum annealer, whereas the QRL-PI needs a range of updates to infer the optimum. The algorithms QEPI and QRL-PI are of comparable complexity in terms of the total number of states \(\mu\). However, it is barely possible to assess their efficiency since they have different additional sources of complexity. In particular, these are the number of bits for binarization in QEPI and the number of samples to estimate value function in QRL-PI, respectively.}

\section*{Code availability}

The source code implementing QEPI, together with the scripts
reproducing all the experiments, is openly available at
\url{https://github.com/enuzhin/QEPI}.

\section*{Acknowledgments}
We acknowledge the use of the supercomputer Zhores~\cite{zhoresSuper}. EEN  acknowledges the support of the Analytical Center (subsidy agreement 000000D730321P5Q0002, Grant No. 70-2021-00145 02.11.2021). DY acknowledges the support from the Russian Science Foundation Project 22-11-00074.

\bibliographystyle{unsrt}
\bibliography{general}

\appendix
\renewcommand{\theequation}{\thesection\arabic{equation}}
\numberwithin{equation}{section}

\section{Time complexity for dense tensors}
\label{sec:dense_time_complexity}

A Markov decision process is P-complete~\cite{littman2013complexity} with the time-to-solution being highly dependent on the problem specifics. In the following, we are to evaluate time complexity of the QEPI in the worst-case scenario. The problem representation in terms of SLE, specified by vector \(b\) and matrix \(A\) in Eqs.~\eqref{b} and \eqref{A} requires
\begin{equation}
    T_{\mathrm{RL}\hookrightarrow\mathrm{SLE}} = \mathcal{O}(\rho\mu^2) + \mathcal{O}(\rho\mu^2) =\mathcal{O}(\rho\mu^2)
\end{equation}
operations in total, where \(\rho\) is the maximal number of distinct rewards possible in conditional (on \(s,a,s'\)) transition, whilst \(\mu\) is the total number of states.

Translation of SLE to a QUBO problem necessitates matrix \(P\) and vector \(p\) as yielded by Eqs.~\eqref{P} and \eqref{p}. A direct computation of the vector \(p\) requires \(n_b\) scaled matrix-vector multiplication operations, whose complexity we estimate as \(\mathcal{O}(\mu^2 + \mu n_b)\). In the latter, we take into account the assumption that we compute \(b^T A\) only once for the first block and reuse the result for the remaining blocks. For the matrix \(P\) we have to perform \(n_b^2\) scaled matrix-matrix multiplications, while this result of the matrix multiplication is reused in follow-up calculations. Thus, the complexity can be estimated as \(\mathcal{O}(\mu^3 + \mu^2n_b^2)\). We herein assume that we use brute-force matrix-matrix multiplications with complexity of \(\mathcal{O}(\mu^3)\). Finally, summing up the complexities of computations of the matrix \(P\) and vector \(p\), we evaluate that of translating SLE to a QUBO problem
\begin{equation}
     T_{\mathrm{SLE}\hookrightarrow\mathrm{QUBO}} = \mathcal{O}(\mu^2 + \mu n_b) + \mathcal{O}(\mu^3 + \mu^2n_b^2) + \mathcal{O}(\mu n_b) =  \mathcal{O}(\mu^3 + \mu^2n_b^2),
\end{equation}
where \( \mathcal{O}(\mu n_b)\) is the complexity of summation, required to compute the matrix \(P+\mathrm{diag}(p)\).

The time complexity of solving a QUBO problem on a quantum annealer heavily depends on problem-specific Hamiltonian~\cite{morita2008mathematical}. We denote this complexity as
\begin{equation}
    T_\mathrm{QA} = \tau_\mathrm{QA},
\end{equation}
which is also sensitive to the problem size. The complexity of solution recovery, according to Eq.~\eqref{bin_enc}, is
\begin{equation}
    T_{R} = \mathcal{O}(\mu n_b),
\end{equation}
and policy update complexity is
\begin{equation}
    T_{\pi} = \mathcal{O}(\alpha\rho\mu^2), 
\end{equation}
where \(\alpha\) is the number of agent actions. 

The final time complexity of one cycle of the algorithm is
\begin{equation}
\begin{split}
    T_\mathrm{update} &= T_{\mathrm{RL}\hookrightarrow\mathrm{SLE}} + T_{\mathrm{SLE}\hookrightarrow \mathrm{QUBO}}  + T_\mathrm{QA}+ T_{R} + T_{\pi}\\
    &= \mathcal{O}(\rho\mu^2) +  \mathcal{O}(\mu^3 + \mu^2n_b^2) +\tau_\mathrm{QA} + \mathcal{O}(\mu n_b) + \mathcal{O}(\alpha\rho\mu^2) =  \mathcal{O}(\mu^3 + \mu^2(n_b^2+\alpha\rho) + \tau_\mathrm{QA}) .
\end{split}
\end{equation}
Upper bound  \(\epsilon\)-optimal convergence time of the policy iteration may be estimated by a constant \cite{littman2013complexity}:
\begin{equation}
    n_\mathrm{PI} = \frac{B + \log(1/\epsilon) + \log(1/(1-\gamma)+1)}{1-\gamma},
\end{equation}
such that number of policy updates required to converge
\begin{equation}
    n^* \le n_\mathrm{PI}.
\end{equation}
This implies an upper bound fixed number of operations for the solution with desired accuracy \(\epsilon\) to the optimal value function. It is supposed that the latter may be expressed as a solution of a linear program task with rational components of no more than \(B\) bits each. In addition, we here use the dominance relation~\cite{littman2013complexity}, which states that policy iteration converges no more slowly than value iteration. Consequently, the upper bound convergence time of the QEPI algorithm may be estimated as
\begin{equation}
    T_\mathrm{QEPI} = \mathcal{O}(\mu^3 + \mu^2(n_b^2+\alpha\rho) + \tau_\mathrm{QA}).
\end{equation}

\section{Space complexity for dense tensors}
\label{sec:dense_space_complexity}

Space complexity, in other words the required memory of a classical computer, does not depend on the number of iterations and annealer specifics. To store the problem-specific transition probabilities \(p(s',r|s,a)\) one has to have \(\mathcal{O}(\alpha\rho\mu^2)\) memory units. The policy \(\pi(s)\) storage requires \(\mathcal{O}(\alpha\mu)\) memory units. And value function \(V(s)\) storage consumes \(\mathcal{O}(\mu)\). To sum up,

\begin{equation}
    M_{I} = \mathcal{O}(\alpha\rho\mu^2) + \mathcal{O}(\alpha\mu) + \mathcal{O}(\mu) = \mathcal{O}(\alpha\rho\mu^2)
\end{equation}
memory units in total have to be used for initialization.

To recast the policy evaluation as an SLE we need to store a matrix \(A\) of size \(\mu^2\) and vector \(b\) of size \(\mu\). This is
\begin{equation}
    M_{\mathrm{RL}\hookrightarrow\mathrm{SLE}} = \mathcal{O}(\mu^2)
\end{equation}
memory units in total.

The translation of SLE to a QUBO problem necessitates storing \(A^TA\) matrix of size \(\mu^2\) for computing the matrix \(P\) of size \(\mu^2n_b^2\) in Eq.~\eqref{P}. To compute a vector \(p\) in Eq.~\eqref{p} we need to store \(b^TA\) which consumes \(\mu\) memory units and storage of vector \(p\) takes \(\mu n_b\). Finally, the QUBO problem is described in terms of the matrix of the size \(\mu^2n_b^2\). In other words,
\begin{equation}
    M_{\mathrm{SLE}\hookrightarrow\mathrm{QUBO}} = \mathcal{O}(\mu^2) + \mathcal{O}(\mu^2n_b^2)+\mathcal{O}(\mu) + \mathcal{O}(\mu n_b) + \mathcal{O}(\mu^2 n_b^2)  =  \mathcal{O}(\mu^2n_b^2) 
\end{equation}
memory units in total.

For restoring the solution we will also have to use no more than
\begin{equation}
    M_R = \mathcal{O}(\mu n_b)
\end{equation}
memory units to receive the QUBO solution and recover it in already allocated memory for value function, as specified by Eq.~\eqref{bin_enc}. The policy update does not require additional memory.
Then, the overall space complexity of the algorithm on classical computers is
\begin{equation}
\begin{split}
    M_\mathrm{QEPI} &= M_I + M_{\mathrm{RL}\hookrightarrow\mathrm{SLE}} + M_{\mathrm{SLE}\hookrightarrow\mathrm{QUBO}} + M_{R}\\
    &= \mathcal{O}(\alpha\rho\mu^2) + \mathcal{O}(\mu^2) + \mathcal{O}(\mu^2n_b^2) +  \mathcal{O}(\mu n_b)
    =  \mathcal{O}(\mu^2(n_b^2 + \alpha\rho)).
\end{split}
\end{equation}

\section{Band tensors}
\label{sec:band_tensors}
Sparse transition tensors are widespread in the domain of RL. The transition is described by the probability \(p(s',r|s,a)\), and if the state space is continuous, then, as a rule, the transitions with non-zero probability are possible to neighboring states \(s'\) and \(s\) only. Here we assume that the transition probability is described by an indexed tensor. 

We refer to the tensor $T$ as a band tensor in \(s\) and \(s'\) indices with a bandwidth of \(k\) if it meets:
\begin{equation}
\begin{split}
         \big( \forall s \in S, \,\, \forall ds_\mathrm{out} \in \{ds_i \in \mathbb{Z} &: ||ds||_\infty > k,s+ds \in S\} \big) :\,\,\,  T(s+ds_\mathrm{out},s) = 0 \\
         &\,\,\,\,\,\,\textrm{and}\\
         \big(\exists s \in S, \,\, \exists ds_\mathrm{in} \in \{ds_i \in \mathbb{Z} &: ||ds||_\infty = k,s+ds \in S\} \big): \,\,\, T(s+ds_\mathrm{in},s) \ne 0,
\end{split}
\end{equation}
where \( ||ds||_\infty = \max_i |ds_i|\) is the $L_\infty$ norm, and \(S\) is a space of multidimensional indexes \(s = \{i_1,...,i_N\}\) and \(s' = \{i_{N+1},...,i_{2N}\}\) of tensor \(T\). In other words, if \(T(s',s) = p(s',r|s,a)\) is transition probability, then \(k\) is the longest distance for which jumps between two states are allowed. 

\section{Time complexity for sparse tensors}
\label{sec:sparse_time_complexity}
Assume that the transition probability is a band tensor in \(s\) and \(s'\) indices with (maximal with respect to the action \(a\) and reward \(r\) dimensions) bandwidth of \(k\) (see Appendix~\ref{sec:band_tensors} for definition). The result of building SLE from known transition probability tensor \(p(s',r|s,a)\) is matrix \(A\) and vector \(b\), see Eqs. \eqref{b} and \eqref{A}. The computation of vector \(b\) leads to
\begin{equation}
    b_i = \sum_{s',r} p(s',r|s_i,\pi(s_i)) \cdot r = \sum_{||ds||_\infty \le k}  \sum_{r} p(s_i + ds,r|s_i,\pi(s_i)) \cdot r,
\end{equation}
where we applied the definition of the band tensor to transition probability \(p(s',r|s_i,\pi(s_i))\) in order to exclude explicitly zero entities from summation.
Since \(||ds||_\infty \le k\) is \(N\)-th dimensional hypercube with the edge length of \(2 k\), where \(N\) is the dimension number of the state vector, the total number of integer vectors \(ds\) in that hypercube is \(a=(2k+1)^N\). Then, the total number of operation to compute  \(b_i\) is  \(\mathcal{O}(a\rho)\) and, consequently, \(b\) requires \(\mathcal{O}(a\rho\mu)\) operations in total.

Similarly, we can compute a number of operations required for matrix A calculation:
\begin{equation}
\label{A_sparse}
    A_{ij} = \delta_{ij} - \gamma  \sum_{r} p(s_j,r|s_i,\pi(s_i)) = \delta_{||s_i-s_j||_\infty \le k} \left[ \delta_{i,j} - \gamma  \sum_{r} p(s_j,r|s_i,\pi(s_i))\right],
\end{equation}
where we highlighted with delta symbol that we do not perform any computation for known in advance zero entities. It follows directly from the definition of band tensor that non-zero probabilities exist for \(s_i\) and \(s_j\) such that \(||s_i-s_j||_\infty \le k\) only. Then, taking into account that each state \(s_i\) has \(a\) neighbors we have overall complexity \(\mathcal{O}(a\rho\mu)\). Summing up the complexities of computation \(A\) and \(b\) we have the total time complexity of policy evaluation problem translation in SLE 

\begin{equation}
    T_{\mathrm{RL}\hookrightarrow\mathrm{SLE}} =\mathcal{O}(a\rho\mu) + \mathcal{O}(a\rho\mu) = \mathcal{O}(a\rho\mu).
\end{equation} 

In order to estimate the complexity of rewriting SLE in terms of QUBO we should again remember that the transition probability tensor \(p(s',r|s,a)\) for a fixed reward \(r\) and action \(a\) is a band tensor with a bandwidth of \(k\). Then, in its tensorial form \(A = A(s,s')\) is also a band tensor with the bandwidth of \(k\). Taking this into account matrix-vector multiplication in Eq.~\eqref{p} for vector \(p\) computation in the form
\begin{equation}
    [b^T A](s) = \sum_{s'} b(s')A(s',s) = \sum_{||ds||_\infty \le k} b(s+ds)A(s+ds,s)
\end{equation}
needs at least \(\mathcal{O}(a\mu)\) operations. This finally gives \(\mathcal{O}(a\mu + \mu n_b)\) operations to compute vector \(p\).

Matrix-matrix multiplication in Eq. \eqref{P} for \(P\) may be described by the following equation:
\begin{equation}
\begin{split}
    [A^T A](s,s') = \sum_{\sigma} A(\sigma,s)A(\sigma,s') &= \sum_{||ds||_\infty \le k} A(s+ds,s)A(s+ds,s')\\
    &= \sum_{||ds||_\infty \le k} \delta_{||s-s'+ds||_\infty \le k} A(s+ds,s)A(s+ds,s').
\end{split}
\end{equation}
Here we utilize the properties of band tensors to exclude zero entities from summation and denote explicitly with delta function known in advance zero terms. Using triangle inequality \(||s-s'||_\infty \le ||s-s'+ds||_\infty + ||ds||_\infty\), we can estimate a lower bound of \( ||s-s'+ds||_\infty \ge ||s-s'||_\infty  - ||ds||_\infty \ge ||s-s'||_\infty  - k \) which guarantees that we do not miss non-zero entities in summation, then
\begin{equation}
\label{ATA_sparse}
\begin{split}
    [A^T A](s,s') &= \sum_{||ds||_\infty \le k} \delta_{||s-s'+ds||_\infty \le k} A(s+ds,s)A(s+ds,s')\\
    &=\sum_{||ds||_\infty \le k} \delta_{||s-s'||_\infty -k \le k} A(s+ds,s)A(s+ds,s')\\
    &= \delta_{||s-s'||_\infty \le 2k} \sum_{||ds||_\infty \le k}  A(s+ds,s)A(s+ds,s').
\end{split}
\end{equation}
Consequently, tensor \([A^T A](s,s')\) is a band tensor with at least bandwidth of \(2k\). Summarizing the number of required operations we need \(\mathcal{O}(a)\) operations in summation for each states pair  \(s\) and \(s'\), but the entities of tensor  \([A^T A](s,s')\) are non-zero for \(c\mu\) elements only due to bandwidth of \(2k\) [we have defined $c=(4k+1)^N$]. Then matrix-matrix multiplication in Eq. \eqref{P} for  \(P\) takes \(\mathcal{O}(\mu ac)\) elementary operations in total. The matrix \(P\), according to Eq. \eqref{P}, has  \(n_b^2\) scaled matrix-matrix multiplications. The complexity of the computation, provided the assumption that we calculate the matrix-matrix multiplication only once, is estimated as \(\mathcal{O}(\mu ac+ \mu cn_b^2)\).

Finally, summing up the complexities of calculation of the matrix \(P\) and vector \(p\), the final complexity of translating SLE to a QUBO problem is
\begin{equation}
\begin{split}
    T_{\mathrm{SLE}\hookrightarrow\mathrm{QUBO}} &= \mathcal{O}(\mu a + \mu n_b) + \mathcal{O}(\mu ac+ \mu cn_b^2) + \mathcal{O}(\mu n_b)\\
    &=  \mathcal{O}(\mu ac+ \mu cn_b^2).
\end{split}
\end{equation}

The time complexity of QUBO solution on a quantum annealer and solution recovery is the same as discussed in Appendix~\ref{sec:dense_time_complexity}. The complexity of annealing is
\begin{equation}
    T_\mathrm{QA} = \tau_\mathrm{QA},
\end{equation}
which is a function of the problem size and the complexity of solution recovery is 
\begin{equation}
    T_{R} = \mathcal{O}(\mu n_b).
\end{equation}
The complexity of the policy update step may be derived from the following equation:
\begin{equation}
    \pi(s) = \arg \max_a\sum_{s',r} p(s',r|s,a)\cdot\{r + \gamma V(s')\} = \arg \max_a\sum_{||ds||_\infty \le k,\,r} p(s +ds,r|s,a)\cdot\{r + \gamma V(s + ds)\},
\end{equation}
which was reduced on condition that the transition probability \(p(s',r|s,a)\) is a band tensor of the bandwidth \(k\). Consequently, the policy update requires 
\begin{equation}
    T_{\pi} = \mathcal{O}(a\alpha\rho\mu)
\end{equation}
elementary operations.

The final time complexity of one cycle of the algorithm for band transition probability tensor is
\begin{equation}
\begin{split}
    T_\mathrm{update} &= T_{\mathrm{RL}\hookrightarrow\mathrm{SLE}} + T_{\mathrm{SLE}\hookrightarrow \mathrm{QUBO}}  + T_\mathrm{QA} + T_{R} + T_{\pi}\\
    &=\mathcal{O}(a\mu\rho) + \mathcal{O}(ac\mu+ c\mu n_b^2) +  \tau_\mathrm{QA} + \mathcal{O}(\mu n_b) + \mathcal{O}(a\alpha\rho\mu)\\
    &= \mathcal{O}(ac\mu+ c\mu n_b^2 +a\alpha\rho\mu + \tau_\mathrm{QA}).
\end{split}
\end{equation}
Following the reasons described in Appendix~\ref{sec:dense_time_complexity} the upper bound convergence of the QEPI algorithm may be estimated based on the time complexity of one cycle of the algorithm as
\begin{equation}
    T_\mathrm{QEPI} = \mathcal{O}(ac\mu+ c\mu n_b^2 + a\alpha\rho\mu + \tau_\mathrm{QA}).
\end{equation}

\section{Space complexity for sparse tensors}
\label{sec:sparse_space_complexity}
To store the problem-specific transition probabilities \(p(s',r|s,a)\) one has to have \(\mathcal{O}(\alpha\rho\mu^2)\) memory units, but due to sparsity of band transition tensor of the bandwidth \(k\) we can reduce this number to \(\mathcal{O}(a\alpha\rho\mu)\) units. The policy \(\pi(s)\) storage requires \(\mathcal{O}\left(\alpha\mu\right)\) memory units. And the value function \(V(s)\) storage needs \(\mathcal{O}(\mu)\). This is
\begin{equation}
    M_{I} = \mathcal{O}(a\alpha\rho\mu) + \mathcal{O}(\alpha\mu) + \mathcal{O}(\mu) = \mathcal{O}(a\alpha\rho\mu)
\end{equation}
memory units in total for initialization.

To recast the policy evaluation as an SLE we need to store matrix \(A\) of the size \(\mu^2\) and vector \(b\) of the size \(\mu\). But due to the sparsity of band transition tensor, this matrix has only \(a\mu\) non-zero elements in agreement with Eq.~\eqref{A_sparse}. This is
\begin{equation}
    M_{\mathrm{RL}\hookrightarrow\mathrm{SLE}} = \mathcal{O}(a\mu)
\end{equation}
memory units in total.

Rewriting SLE in terms of a QUBO problem requires storing \(A^TA\) of the size \(\mu^2\). This matrix is sparse, as specified by Eq.~\eqref{ATA_sparse} and in its tensorial form this is a band tensor of the bandwidth \(2k\). To store we need \(c\mu\) memory units.  This matrix is of use to compute \(P\) of the size \(\mu^2 n_b^2\) following Eq.~\eqref{P}. Taking into account sparsity of \(A^TA\), storing matrix \(P\) costs \(c\mu n_b^2 \) memory units.  To compute the vector \(p\) in Eq.~ \eqref{p} we need to store \(b^TA\) which needs \(\mu\) memory units, while storing this vector results in \(\mu n_b \). The final weight matrix that describes the whole QUBO problem is of the size \(\mu^2 n_b^2\) but with the same sparsity properties as matrix \(P\),  since it is computed by addition value of \(p\) to elements corresponding to identical states. Consequently, it costs \(c\mu n_b^2\) memory units to store. This is
\begin{equation}
        M_{\mathrm{SLE}\hookrightarrow\mathrm{QUBO}}= \mathcal{O}(c\mu) + \mathcal{O}(c\mu n_b^2) + \mathcal{O}(\mu)+ \mathcal{O}(\mu n_b) + \mathcal{O}(c\mu n_b^2) =  \mathcal{O}(c\mu n_b^2) 
\end{equation}
memory units in total.

Solution recovery needs no more than
\begin{equation}
    M_R = \mathcal{O}(\mu n_b)
\end{equation}
memory units to receive the QUBO solution and restore it to already allocated memory for the value function. The policy update does not require additional memory.

Finally, the overall space complexity of the algorithm on classical computers is
\begin{equation}
\begin{split}
    M_\mathrm{QEPI} &= M_I + M_{\mathrm{RL}\hookrightarrow\mathrm{SLE}} + M_{\mathrm{SLE} \hookrightarrow\mathrm{QUBO}} + M_{R} \\
    &=\mathcal{O}(a\alpha\rho\mu) + \mathcal{O}(a\mu) + \mathcal{O}(c\mu n_b^2)  +  \mathcal{O}(\mu n_b) = \mathcal{O}(c\mu n_b^2 + a\alpha\rho\mu).
\end{split}
\end{equation}

\section{Quantum annealer}
\label{sec:qa_requirenments}
The main quantity which describes the quantum annealer performance is the number of qubits allowed to specify the problem. In order to apply the QEPI algorithm we need to estimate the minimum number of qubits required for the algorithm. The QUBO problem, as specified by Eqs.~\eqref{bin_enc} and \eqref{obj}, is formulated
in terms of \(n_b\) binary variables per state, which amounts to \(\mu n_b\) variables
in total. In the annealing paradigm each binary variable is represented by a single
qubit, in contrast to the amplitude encoding employed by gate-based
solvers~\cite{harrow2009quantum,chakraborty2018power}. Hence, for the particular RL
problem we need at least 
\begin{equation}
    N_\mathrm{qubits} = \mu n_b
\end{equation}
qubits, where \(\mu\) is the total number of states and \(n_b\) is the number of bits
encoding the discrete level of the value function at a given state.
 
Still, some quantum computers can not implement full-scaled inter-qubit connections~\cite{shin2014quantum}. If the transition probability \(p(s',r|s,a)\) is a band tensor of the bandwidth of \(k\) in state indices \(s\) and \(s'\), as discussed in Appendix~\ref{sec:band_tensors}, the resulting matrix of the QUBO problem has \(c\mu n_b^2\) non-zero elements in \(N\)-dimensional state space. The proof to this statement is provided in Appendix~\ref{sec:sparse_space_complexity}. Consequently, the resulting matrix of the QUBO problem has sparsity
\begin{equation}
    \textrm{Sparsity}(Q) \ge 1-\frac{c}{\mu},
\end{equation}
which is a common case for deterministic optimal control problems augmented with
stochastic state discretization.
\end{document}